\documentclass[journal]{IEEEtran}

\usepackage{cite}

\ifCLASSINFOpdf
  \usepackage[pdftex]{graphicx}
  \graphicspath{{../pdf/}{../jpeg/}}
  \DeclareGraphicsExtensions{.pdf,.jpeg,.png}
\else
   \usepackage[dvips]{graphicx}
   \graphicspath{{../eps/}}
   \DeclareGraphicsExtensions{.eps}
\fi

\usepackage[cmex10]{amsmath}

\usepackage{algorithmic}

\usepackage{array}
\usepackage{mdwmath}
\usepackage{mdwtab}
\usepackage{eqparbox}

\usepackage{multirow}
\usepackage{xtab}
\usepackage{booktabs}
\usepackage{tabularx}
\usepackage{rotating}
\usepackage{longtable}
\usepackage{hyperref}

\usepackage{url}
\usepackage{subfigure}
\usepackage{graphicx}

\usepackage{balance}
\usepackage{comment}
\usepackage{subfigure}
\usepackage{graphicx}
\usepackage{epsfig}
\usepackage{url}
\usepackage{amsfonts}
\usepackage{amssymb}
\usepackage{mdwmath}
\usepackage{mdwtab}
\usepackage{cite}
\usepackage{booktabs}
\usepackage{tabularx}
\usepackage{multirow} 
\usepackage[]{caption}
\usepackage{amsmath}

\usepackage{xcolor}

\begin{document}
\title{CheepSync: A Time Synchronization Service for Resource Constrained Bluetooth LE Advertisers}
\author{
\IEEEauthorblockN{Sabarish Sridhar\IEEEauthorrefmark{1}, Prasant Misra\IEEEauthorrefmark{2}, Jay Warrior\IEEEauthorrefmark{2}}\\
\IEEEauthorblockA{\IEEEauthorrefmark{1}M.S. Ramaiah Institute of Technology, Bangalore, India\\
\IEEEauthorrefmark{2}Robert Bosch Center for Cyber Physical Systems, Indian Institute of Science, Bangalore, India\\
Email: sabarishvs111@gmail.com\IEEEauthorrefmark{1} \{prasant.misra, jay.warrior\}@rbccp.org\IEEEauthorrefmark{2}}
}

\IEEEcompsoctitleabstractindextext{%
\begin{abstract}
Clock synchronization is highly desirable in distributed systems, including many applications in the Internet of Things and Humans (IoTH).
It improves the efficiency, modularity and scalability of the system; and optimizes use of event triggers. 
For IoTH, Bluetooth Low Energy (BLE) - a subset of the recent Bluetooth v$4.0$ stack - provides a low-power and loosely coupled mechanism for sensor data collection with ubiquitous units (e.g., smartphones and tablets) carried by humans.
This fundamental design paradigm of BLE is enabled by a range of \emph{broadcast advertising} modes.
While its operational benefits are numerous, the \emph{lack} of a common time reference in the broadcast mode of BLE has been a fundamental limitation.
This paper presents and describes \emph{Cheep}Sync
: a time synchronization service for BLE advertisers, especially tailored for applications requiring high time precision on resource constrained BLE platforms.
Designed on top of the existing Bluetooth v$4.0$ standard, the \emph{Cheep}Sync framework utilizes low-level timestamping and comprehensive error compensation mechanisms for overcoming uncertainties in message transmission, clock drift and other system specific constraints.
\emph{Cheep}Sync was implemented on custom designed nRF$24$\emph{Cheep} beacon platforms (as broadcasters) and commercial off-the-shelf Android ported smartphones (as passive listeners).
We demonstrate the efficacy of \emph{Cheep}Sync by numerous empirical evaluations in a variety of experimental setups; and show that its average (single-hop) time synchronization accuracy is in the $10$\,$\mu$s range.
\end{abstract}
}

\maketitle
\IEEEdisplaynotcompsoctitleabstractindextext
\IEEEpeerreviewmaketitle










\vspace{-2mm}
\section{Introduction} \label{sec:introduction}

A common time reference is an important requirement for distributed systems.
The accurate knowledge of time is primeval for concurrency control, data consistency/ordering, security, communication protocol management, etc., in wireless networks, network control and sensor networks; without which the system performance, Quality of Service (QoS), and safety would get \emph{severely} affected.
Time synchronization across distributed clocks is an area with a rich history.
It includes a whole gamut of solutions covering the notion of ordering of events with virtual clocks\cite{Lamport1978:TCO} to relative/absolute synchronization\cite{Maroti2004:FTSP,Elson2002:RBS,Ganeriwal2003:TPSN,Ferrari2011:glossy,Mills1991:NTP,Gusella1989:tempo,Cristian1989:probtimesync,Kopetz1987:CSD,Rowe2009:LCS} for platforms that are resource savvy to constrained.
In this regard, developing a common time service for distributed systems consisting of constrained devices is particularly challenging.
\newline
\indent
In the recent past, a new class of constrained platforms based on Bluetooth Low Energy (BLE)\cite{BLE} have emerged.
BLE is different from other wireless technologies because it combines a standardized communication technology designed for low-power systems,
and a new sensor-based data collection framework.
It also offers easy integration with most handheld devices (such as smartphones and tablets), something that traditional wireless sensor networks (WSN) are still working towards. 
BLE has inherited several technical features from Classic Bluetooth that provide for robust, reliable connections. 
However, the most significant difference is its \emph{asymmetric} design.
While the standard mode of operation is typically based on a master device connected to a number of slave devices, it offers a new feature in the form of an \emph{`advertisement'} (configurable through the broadcast/peripheral mode of BLE). 
This new mode offers unidirectional communication between two or more LE devices using advertising events, thereby achieving a communication solution without entering into a bonded connection (as required by Classic Bluetooth devices).
Such a loose coupled manner of data transfer is undoubtedly more energy efficient; but also unearths other limitations. 
For example, the broadcast mode of communication \emph{does not} provision for a time synchronization service; even though this feature is included in the solution stack, but can \emph{only} be availed upon pairing.
\vspace{2mm}
\newline
\noindent
\textbf{Challenges.}
BLE provides a range of broadcast advertising modes, of which the \emph{most} energy efficient is the non-connectable undirected advertising mode (\texttt{ADV}\_\texttt{NONCONN}$\_$\texttt{IND}): a transmit-only, broadcaster mode \emph{without} any listen window.
Establishing time synchronization in this mode of BLE operation is \emph{challenging} due to many reasons.
\emph{First}, the traditional techniques of message passing among different elements is not supported by this network architecture (i.e., Bluetooth v$4.0$ specific); and as a result, timing uncertainties cannot be compensated by \emph{exchanging} time stamped packets, or `pings' between nodes.
\emph{Second}, devices receiving advertisement data (such as smartphones) have high functional asymmetry compared to the BLE broadcast units.
They typically run on a multithreaded and multitasking operating systems (OS) (such as Android) where the \emph{measure} of system latencies and their associated uncertainties can be many orders of magnitude higher than those at the transmitter end.
\emph{Third}, low-level timestamping on such multifunctional receiver devices can be performed to a \emph{certain} limit and is subject to system restrictions of the underlying firmware.
Therefore, motivated by the need to overcome the above limitations and yet be able to establish a common time reference across resourced constrained BLE devices operating in the \texttt{ADV}\_\texttt{NONCONN}$\_$\texttt{IND} mode, we propose \emph{Cheep}Sync.
\vspace{2mm}
\newline
\noindent
\textbf{Contributions and Road-map.}
Due to the architectural constraints, the key ideas of \emph{Cheep}Sync are: ($1$) not to synchronize the nodes in the network, but to make the devices that use these nodes synchronize; and ($2$) piggyback on the device mobility aspect (as they are inevitably carried by people) and use them as `synchronization mules' for the `broadcast' system.
Thus, it offers a flexible piggyback design wherein running the time service does not require data transactions to be temporarily suspended.
Therefore, time synchronization with \emph{Cheep}Sync is highly implicit rather than explicit.
Since \emph{Cheep}Sync rides on the BLE broadcast framework, it is scalable to the point that the framework has to offer.
\newline
\indent
In this article, we provide a synopsis of the BLE broadcast mode and Bluedroid (the default Android Bluetooth stack) along with system details of building a BLE fakery platform in the next section.
It is followed by a detailed design and analysis of the \emph{Cheep}Sync architecture and performance that is able to achieve an average time synchronization accuracy in the range of $10$\,$\mu$s.
The final sections provide a concise background of existing work in this related field, and concludes with a summary of the areas covered in the article. 

\section{System Overview} \label{sec:system_overview}

\begin{figure*}[t]
\label{fig:ble-platform}
\begin{center}
\subfigure[Platform overview]{\label{fig:}\includegraphics[width=3in]{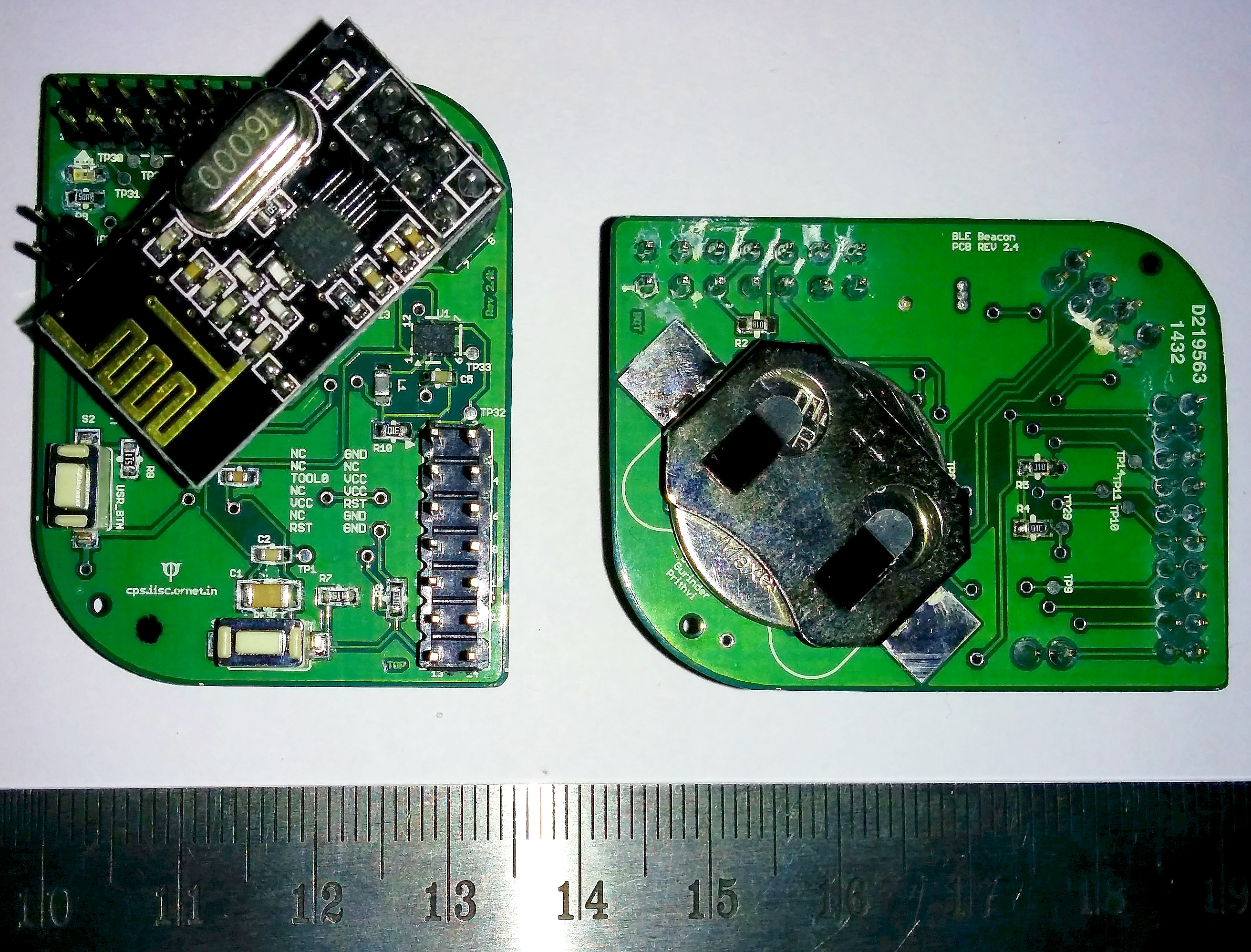}} \hspace{1cm}
\subfigure[Functional representation of the major platform components]{\label{fig:}\includegraphics[width=3.3in]{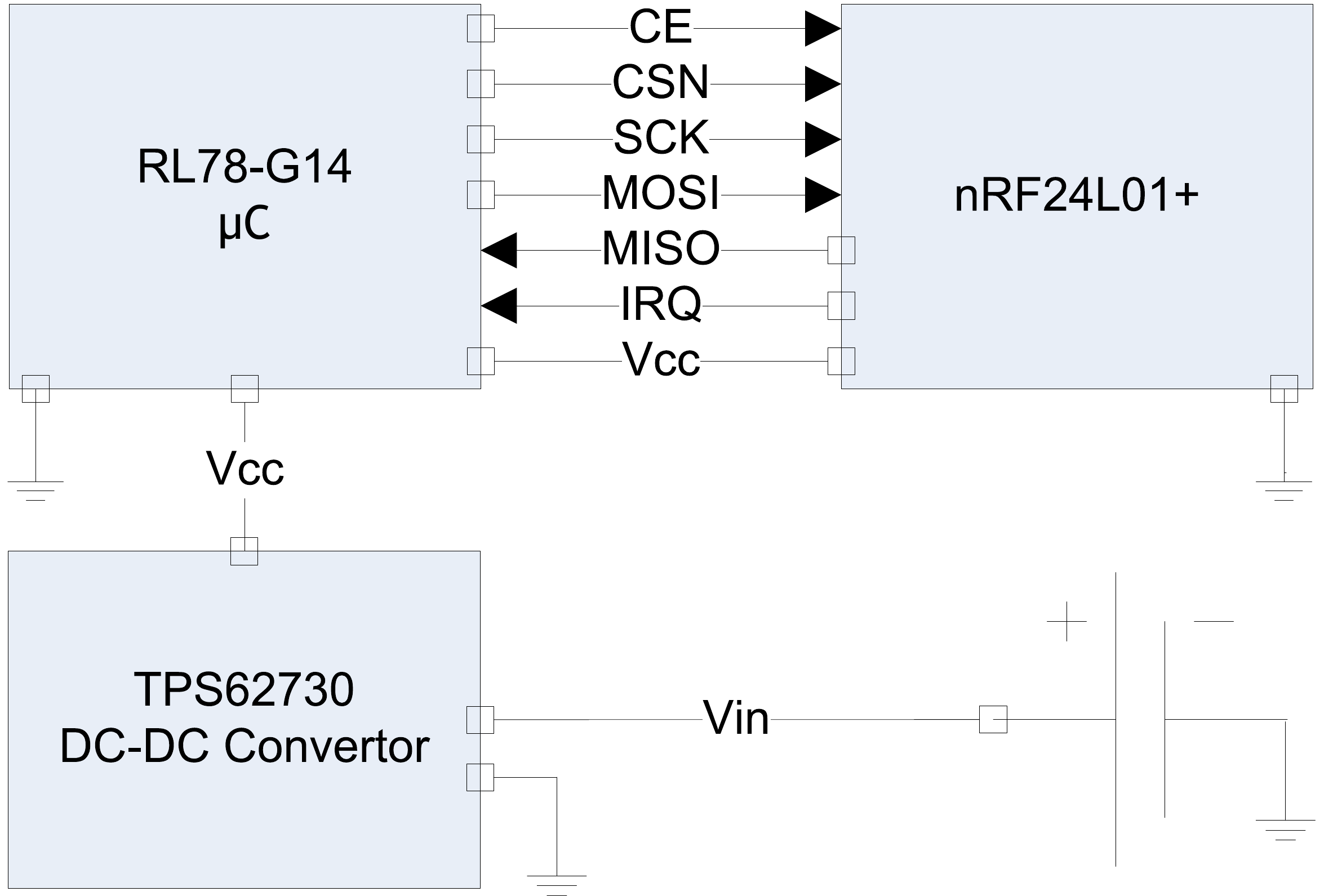}} 
\end{center}
\caption{\textbf{nRF$24$\emph{Cheep}: custom designed BLE Beacon platform.} \emph{The major components include:} RL$78$-G$14$\cite{RenesasRL78G14} \emph{microcontroller}, NRF$24$L$01$+\cite{nRF2401} $2.4$\,GHz \emph{RF transceiver}, TPS$62730$ \emph{synchronous step-down DC-DC converter, and} CR$2032$ \emph{battery holder.}}
\end{figure*}

The system is composed of \emph{two} units: beacon and control.
The \emph{beacon unit} consists of resource constrained sensing tags that are deployed in the region of interest.
They are responsible for measurement of simple physical parameters, and disseminating that information through undirected BLE broadcasts. 
The \emph{control unit} consists of a resourceful gateway device capable of listening and receiving broadcast data contained in BLE advertisements.  
For our application requirements, the beacon unit is a custom designed nRF$24$\emph{Cheep} BLE platform, and the control unit is an Android v$4.4.4$ smartphone that uses Bluedroid (the default Android Bluetooth stack).
\newline
\indent
To ground our discussion, we first provide an overview of BLE in this section, followed by a detailed explanation of the nRF$24$\emph{Cheep} platform and the Android Bluedroid stack.
\vspace{-4mm}
\subsection{BLE Synopsis}
BLE is a subset of the Bluetooth v$4.0$ stack, and is designed for \emph{low-power} applications interested in knowing the \emph{state} of the physical world. 
It is, therefore, not suitable for high data rate applications, which are better served by the existing Bluetooth $2.0$ EDR or Bluetooth $3.0$ HS protocols.
BLE enabled systems are inherently asymmetric in-nature with large numbers of simple, resource constrained peripherals devices (that interact with the physical space); and few complex, resource savvy central devices (that act as gateways to the cyber space).
\newline
\indent
BLE operates in the unlicensed $2.4$\,GHz ISM band.
It defines $40$ channels (numbered $0$-$39$), each separated by $2$\,MHz.
\emph{Three} of these channels ($37$, $38$, $39$) are reserved for advertising and are used for device discovery, connection establishment and broadcast (one-way) communication; while the remaining channels are dedicated for bi-directional communication between connected devices.
BLE ensures reliable coexistence with other competing RF technologies (especially WiFi) operating in the ISM band by using non-overlapping frequencies for the three advertisement channels, while it adopts an adaptive frequency hopping mechanism for the data channels.
\newline
\indent
In the broadcast mode of communication, as part of an advertisement event, the advertiser (i.e., the peripheral device) sequentially transmits an advertisement packet on each advertisement channel within a specified time period.
Broadcast packet transmissions are kept short with only $31$\,bytes available for application payload, in addition to other necessary fields such as the preamble, header, Medium Access Control (MAC) address and checksum.
Such packets are received by any central device that is passively listening to such broadcasts.
For bidirectional and unconstrained data transfer, the respective devices need to be paired and connected. 
The data exchange in this mode is enabled by the Generic Access Profile (GAP), a light-weight client-server mechanism that also supports predefined or user defined Generic Attribute (GATT) services.
\newline
\indent
BLE defines \emph{four} type of broadcast mechanisms: ($1$) connectable undirected advertising \texttt{ADV}\_\texttt{IND}, ($2$) connectable directed advertising \texttt{ADV}\_\texttt{DIRECT}$\_$\texttt{IND}, ($3$) non-connectable undirected advertising \texttt{ADV}\_\texttt{NONCONN}$\_$\texttt{IND}, and ($4$) scannable undirected advertising \texttt{ADV}\_\texttt{SCAN}$\_$\texttt{IND}.
Of these, \texttt{ADV}\_\texttt{NONCONN}$\_$\texttt{IND} is the singular transmit-only broadcast mode without any scope for listening back.
Thus, it facilities BLE-enabled platforms to much more constrained in terms of energy and communication capabilities than the current state-of-the-art.
\newline
\indent
In the following subsection, we describe our experiences in building a custom BLE beacon platform that uses BLE fakery over a general purpose radio, a tool for conducting research in this direction.  
\subsection{nRF24Cheep BLE Beacon Platform}
The custom designed beacon platform, nRF$24$\emph{Cheep}, consists of a: RL$78$-G$14$\cite{RenesasRL78G14} microcontroller, nRF$24$L$01$+\cite{nRF2401} $2.4$\,GHz RF transceiver with an embedded baseband protocol engine (Enhanced ShockBurst$^{\texttt{TM}}$), TPS$62730$ synchronous step-down DC-DC converter, and CR$2032$ battery holder.
The microcontroller has a RL$78$ core with $16$-$512$\,KB flash memory, and operates at a maximum clock speed of $64$\,MHz with a high precision ($\pm1$\%) on-chip oscillator. 
It also provides a range of \emph{interval timers} for different application requirements.
The nRF$24$L$01$+, although not strictly compatible with BLE, can be made to operate as a BLE transmitter by configuring the transceiver settings\footnote{The necessary RF configurations for BLE compliance can be obtained from \cite{ble-fakery}.}.
This allows BLE listeners/scanners to view the nRF$24$L$01$+ as a BLE device and decode its advertisement data.
The nRF$24$L$01$+ interfaces with the application controller (RL$78$-G$14$) over a high speed Serial Peripheral Interface (SPI) bus.
Enhanced ShockBurst$^{\texttt{TM}}$, designed to handle all the high speed link layer operations, is based on packet communication and supports $1$ to $32$\,bytes of dynamic payload length.
Data flow between the radio front end and the microcontroller is through internal FIFOs.
\vspace{2mm}
\newline
\noindent
\textbf{nRF$24$L$01$ Hardware Interface:} 
The nRF$24$L$01$ module has the following \emph{eight} interfacing pins, of which four are SPI related: CSN, SCK, MISO, MOSI; and the remaining ones are: Vcc, GND, IRQ, CE. 
\newline
\indent
The SPI interface is used for \emph{data} transmission and reception between the radio (slave) and the microcontroller (master).
The active-low CSN (chip select not) is the enabler pin for the SPI bus, while the SCK pin is its serial clock.
CSN is normally kept high except when there is command/data exchange between radio and the microcontroller.
MOSI (or Master Out, Slave In) is the pin on which the master communicates with the slave, and vice versa for the MISO (or  Master In, Slave Out) pin.
\newline
\indent
CE is used to \emph{control} data transmission and reception in TX and RX modes, respectively.
In TX mode, CE is always held low except when the packet has to be transmitted; and is done by loading the TX FIFO and then toggling the CE pin.
IRQ is the interrupt pin, and can be used to assert three internal interrupts, \emph{viz.}: data received, data transmitted, and maximum number of transmit retries reached.
\subsection{Android BlueDroid Stack} 
BlueDroid is the default Bluetooth stack of Android, and consists of two layers.
The Bluetooth Embedded System (BTE) layer holds the core Bluetooth functionality, while communication with Android framework is handled by  the Bluetooth Application Layer (BTA).
A Bluetooth system service communicates with the Bluetooth stack through Java Native Interface (JNI) and with applications through Binder Inter-process Communication (IPC).
The android application code (at the framework level) utilizes the APIs of \texttt{android.bluetooth} to interact with the bluetooth hardware that internally, through the Binder IPC mechanism, calls the Bluetooth process (both the Bluetooth service and profiles).
The packaged android app uses JNI to call into the hardware abstraction layer (HAL) and also receive callbacks.
Located in: \texttt{external/bluetooth/bluedroid}, the default Bluetooth stack implements the generic Bluetooth HAL as well as customizes it with extensions and configuration changes.

\section{CheepSync:\\ Time Synchronization Protocol} \label{sec:cheepsync}

The basic \emph{Cheep}Sync mechanism uses the broadcaster mode of BLE to transmit a single advertisement packet from the beacon (transmitter) unit to the control (receiver) unit.
The broadcasted message contains: (i) the transmitter's current timestamp value, which is a counter field that increments every $interval$\,ms, and is the estimated local time at the transmission of the advertisement packet; and (ii) the aggregate delay incurred during the transmission of the previous packet. 
On message reception, the receiver obtains the corresponding `wall' clock time expressing time (in nanoseconds) since the epoch.
In principle, \emph{two} broadcast packets provide a \emph{synchronization} point between the transmitter and the receiver\footnote{In Section~\ref{sec:sources_error}, we explain how the determinism of time delays on the nRF$24$\emph{Cheep} beacon platform benefits our implementation. Therefore, for our platform of choice, only a \emph{single} broadcast packet (instead of two) is apt for reliably synchronizing the transmitter and the receiver.}.
The difference between the local and `wall' clock time of a synchronization point estimates the clock offset of the transmitter.
\newline
\indent
The counter value $ts$\_$counter$ is needed to estimate the time elapsed on the beacon unit.
If the underlying platform uses a $b$\,bit field to store the timestamp value and the timer fires every $interval$\,ms, then the total time that can be represented by the timestamp field is ($2^{b}*interval$)\,ms.
Therefore, the configuration of $b$ = $24$\,bits and $interval$ = $100$\,ms can be used to make the timestamp value work for approximately $19$\,days without rolling over\footnote{Our system design is based on the assumption that there must be a control unit that passes by the beacon units, atleast once over a $19$ day period. However, under overriding circumstances, the system can determine the number of counter rollovers.}.
\newline
\indent
In the next subsection, we discuss the general uncertainties associated with RF message delivery and then converge to our specific use case.
It is then followed by a detailed explanation of the structure of the BLE advertisement packet with specifics into the packet restructuring as per the nRF$24$L$01$+ Enhanced ShockBurst$^{\texttt{TM}}$ protocol engine.


\subsection{Sources of Time Synchronization Error} \label{sec:sources_error}

Time synchronization is affected by nondeterministic delays (that are many orders of magnitude greater than the required time accuracy), often arising from random events during the process of radio message delivery.
We shall use the following error decomposition model\cite{Kopetz1987:CSD,Horauer2001:psync,Ganeriwal2003:TPSN,Elson2002:RBS,Maroti2004:FTSP} to better understand the sources of latency; and modify it according to the specifics of the platform and radio-of-interest.
\vspace{0.5mm}
\newline
\noindent
$\bullet$ \textbf{Send Time.} It is the time spent by the transmitter to assemble the message and trigger the send request to the MAC layer.
It is, therefore, a function of the processor load and the system call overhead of the respective OS ported on the transmitter platform.
nRF$24$\emph{Cheep} does not have an OS port, and makes direct system calls to the underlying hardware without any (potential) soft routing.
This enables more user control over different system modules (albeit, increased complexity), and therefore, helps to reduce delays that are typically nondeterministic and were previously difficult to calibrate. 
\vspace{1mm}
\newline
\noindent
$\bullet$ \textbf{Access Time.} It is the delay incurred waiting for access to the transmit channel up to the point when transmission begins, and is specific to the MAC protocol in use.
It is considered the least deterministic part of the message delivery system.
BLE \emph{does} provision for a (TDMA/FDMA) MAC, but it is only operational in connection mode.
Access control rules have not been defined for the BLE broadcast mode of communication; and so, packets get pushed on to the physical channel as and when they are flagged for transmission.
\vspace{1mm}
\newline
\noindent
$\bullet$ \textbf{Transmission Time.} A function of the length of the message and the radio speed, it is the time taken by the transmitter to transmit the message and is a deterministic component.
\vspace{1mm}
\newline
\noindent
$\bullet$ \textbf{Propagation Time.} Once the message has left the transmitter, it is the time needed to transit to the receiver. 
For many application requirements (wherein the channel length is under $300$\,m), this delay is highly deterministic and less than $1$\,$\mu$s.
\vspace{1mm}
\newline
\noindent
$\bullet$ \textbf{Reception Time.} It is the time taken by the receiver to receive the message.
In our case, it is the most nondeterministic part of the message delivery mechanism as the receivers are Android ported smartphones that run multiple tasks and process threads at the same time.
Therefore, depending on the system state, a BLE packet can take several milliseconds to propagate to the application layer.
\vspace{1mm}
\newline
\noindent
$\bullet$ \textbf{Receive Time.} It is the time required to process and notify the incoming message to the receiver application. 
\newline
\indent
We perform low-level timestamping, both at the transmitter and receiver end, to overcome the above-stated uncertainties in a message transaction; the details of which are discussed subsequent to the BLE advertisement packet format that follows next.

%
%
\subsection{Advertisement Packet Format}
There is a single format for a BLE (advertisement or data) \texttt{Packet}, and it consists of the following \emph{four} fields:
\begin{enumerate}
	\item \texttt{Preamble} ($1$\, octet) 
	\item \texttt{Access Address} ($4$\, octets)  
	\item \texttt{Protocol Data Unit} (PDU: conventionally\footnote{By conventional/standard, we mean the guidelines provided in the Bluetooth core specification v$4$.} $2$-$39$\,octets, but limited to $2$-$32$\,octets in nRF Enhanced ShockBurst$^{\texttt{TM}}$ packet format) 
	\item \texttt{Cyclic Redundancy Check} (CRC, $3$\,octets) 
\end{enumerate}
\noindent
As per the core specifications of an advertisement packet, the $8$\,bit preamble and the $32$\,bit access address were set to \texttt{10101010b} and \texttt{0x8E89BED6} respectively.
The preamble is used in the receiver to perform frequency synchronization, symbol timing estimation, and automatic gain control training.
A $24$\,bit CRC is appended to the end of every packet, and is calculated over the PDU.
It is important to note that some fields in the packet definition, marked as RFU, are reserved for future use; and are set to zero on transmission and ignored upon receipt.
Depending on the PDU size, a BLE advertisement packet length could vary from $10$ to $40$ octets in the nRF Enhanced ShockBurst$^{\texttt{TM}}$ mode (as opposed to the standard $10$ to $47$\,octets payload).
The \emph{advertisement} channel PDU has a $16$\,bit \texttt{Header} and a variable size \texttt{Payload}.
\vspace{1mm}
\newline
\noindent
\textbf{Header:} The \texttt{Header} consists of the following \emph{six} fields spanning over $2$\,octets:
\begin{enumerate}
	\item \texttt{PDU type} ($4$\,bits) 
	\item \texttt{RFU} ($2$\,bits)  
	\item \texttt{TxAdd} ($1$\,bit) 
	\item \texttt{RxAdd} ($1$\,bit)  
	\item \texttt{Length} ($6$\,bit)  
	\item \texttt{RFU} ($2$\,bit) 
\end{enumerate}
\noindent
The \texttt{PDU type} was set to \texttt{ADV}\_\texttt{NONCONN}$\_$\texttt{IND} (\texttt{0010b}) for \emph{transmitting} non-connectable undirected advertising events.
The following \texttt{RFU}, \texttt{TxAdd} and \texttt{RxAdd} fields were not used, and hence, were set to zero.
The payload size is indicated by the \texttt{Length} field, and it can vary between $6$ to $30$\,octets (instead of the standard $37$\,octets).
\vspace{1mm}
\newline
\noindent
\textbf{Payload:} The \texttt{Payload} for the \texttt{ADV}\_\texttt{NONCONN}$\_$\texttt{IND} PDU consists of the following \emph{two} fields:
\begin{enumerate}
	\item \texttt{AdvA} ($6$\,octets) 
	\item \texttt{AdvData} ($0$-$24$\,octets, instead of the standard $0$-$31$\,octets) 
\end{enumerate}
\noindent
The \texttt{AdvA} field holds the device address of the advertiser, which can either be public (if \texttt{TxAdd} = $0$) or random (if \texttt{TxAdd} = $1$).
The \texttt{AdvData} field contains the advertisement data, and it consists of \emph{two} logical parts: \texttt{significant} and \texttt{non-significant}.
The \texttt{significant} part contains a sequence of \texttt{AD} structures.
\emph{Each} \texttt{AD} structure consists of the following \emph{two} fields to populate a separate item of user data:
\begin{enumerate}
	\item \texttt{Length} ($1$\,octet) 
	\item \texttt{Data} (\texttt{Length} octets)	
	\begin{itemize}
		\item \texttt{AD type} ($1$\,octet) 
		\item \texttt{AD data} (\texttt{Length-1}\,octets)
	\end{itemize}
\end{enumerate}
The \texttt{non-significant} part extends the Advertising data to the remaining octets and contains all zeroes.
For our implementation, we use the \texttt{AD} structures (defined in the core specification) presented in Table~\ref{tab:adtypes}.
\begin{table}[h]
\begin{center}
\caption{Utilized AD Types}
\begin{small}
\begin{tabular}{p{2.5cm} p{5cm}}
	\toprule
	\textbf{Field} & \textbf{Value} \\
	\midrule
	\texttt{Length} & \texttt{$2$\,octets}\\
	\texttt{AD type} & \texttt{Flags}\\
	\texttt{AD data} & \texttt{General Discoverable Mode}\\
	\midrule
	\texttt{Length} & \texttt{$6$\,octets}\\
	\texttt{AD type} & \texttt{Local Name (Shortened)}\\
	\texttt{AD data} & \texttt{-} \\
	\midrule
	\texttt{Length} & \texttt{$4$\,octets}\\
	\texttt{AD type} & \texttt{Manufacturer Specific Data}\\
	\texttt{AD data} & \texttt{Timestamp, Transmit Time Delay (previous packet)} \\
\bottomrule
\label{tab:adtypes}
\end{tabular}
\end{small}
\end{center}
\end{table}
\vspace{1mm}
\newline
\noindent
The AD type \texttt{Flags} sets the discoverability preference of the device, and the \texttt{General Discoverable Mode} makes it detectable unconditionally.
The \texttt{Local Name (Shortened)} AD type sets the short user-readable name of the device. 
The \texttt{Manufacturer Specific Data} AD Type is a generic, freely formattable data field, and includes the $24$\,bit timestamp value, and a $8$\,bit transmit time delay (of the previous packet). 

\subsection{Timestamping@Transmitter} \label{sec:transmitter_timestamping}
\begin{figure}[t]
\begin{center}
\includegraphics[width=3.6in]{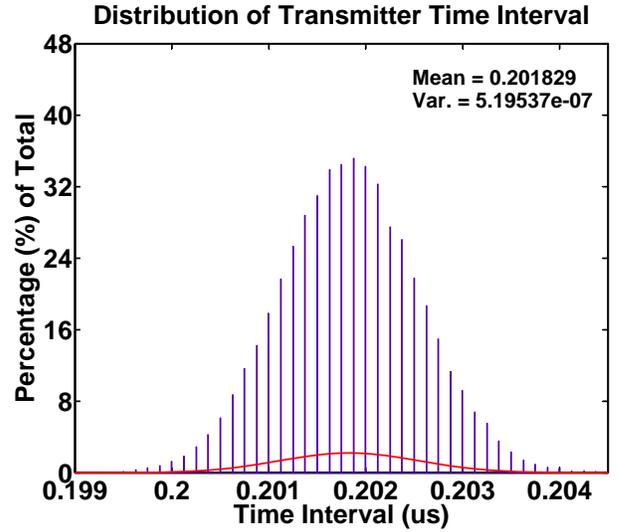}
\end{center}
\vspace{5mm}
\caption{\textbf{Transmitter side time stamping characteristics.} \emph{Histogram showing the distribution of transmitter time interval recorded for $35000$ broadcast packets, grouped into $1$\,$\mu$s buckets. The curve is a plot of the best-fit Gaussian parameters with $\mu$=$0.201828$\,$\mu$s and $\sigma^2$=$5.19537e-07$\,$\mu$s.}}
\label{fig:txts}
\end{figure}

On the nRF$24$\emph{Cheep} platform, an advertisement packet is transmitted by loading the TX FIFO and pulling the \texttt{CE} pin to a high state.
One of the fields that is pushed into the SPI buffer is the current timestamp value.
A second timestamp is also recorded once the \texttt{TX$\_$DS} interrupt is seen by the microcontroller (i.e., when the radio's \texttt{IRQ} pin is pulled down low), signaling the success of the transmit event.
The difference between these two timestamps provides an accurate estimate of the time delays incurred due to send, access, and transmission; and this information is encapsulated into the next advertisement packet.
As discussed in Section~\ref{sec:sources_error}, due to the high determinism of send, access, and transmission time delays on the nRF$24$\emph{Cheep} beacon platform, our implementation uses a \emph{single} broadcast packet (instead of two) to reliably synchronize the transmitter and the receiver.
\newline
\indent
The timestamping characteristics at the transmitter end is shown in Fig.~\ref{fig:txts}, and is depicted as a histogram showing the distribution of the transmitting time interval recorded for $35000$ broadcast packets.
The distribution appears Gaussian with a best fit parameters of $\mu$=$0.201829$\,$\mu$s and a minuscule $\sigma^2$=$5.19537e-07$\,$\mu$s.
This latency characterization supports the determinism of our approach on the transmitter end.

\begin{figure*}[t]
\begin{center}
\subfigure[Before correction: $\mu$=2.62\,ms, $\sigma^2$=6.37312\,$\mu$s, CDF($95$\%)=$8.2$\,ms; After correction: $\mu$=$0.0667$\,ms, $\sigma^2$=3.89512e-02\,$\mu$s, CDF($95$\%)=$0.64$\,ms]{\label{fig:clockdrifta}\includegraphics[width=3.3in]{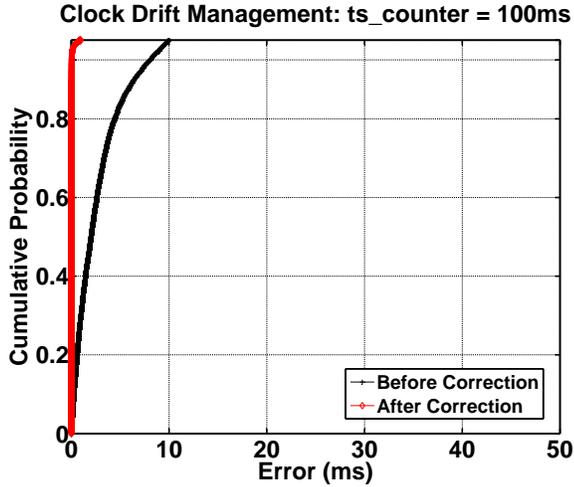}}\hspace{0.7cm}
\subfigure[Before correction: $\mu$=2.3\,ms, CDF($95$\%)=$25$\,ms; After correction: $\mu$=0.0605\,ms, CDF($95$\%)=$2$\,ms]{\label{fig:clockdriftb}\includegraphics[width=3.3in]{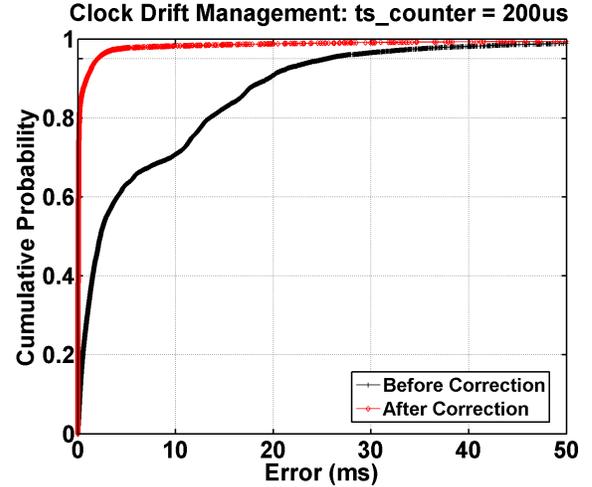}}
\end{center}
\caption{\textbf{Clock Drift Management.} \emph{There is one order improvement in synchronization accuracy after compensating for clock drift, both in the average and $95$\% probability level. The accuracy and stability of the interval timer when set to $100$\,ms is significantly better than the case when it is configured to $200$\,$\mu$s.}}
\label{fig:clockdrift}
\end{figure*}

\subsection{Timestamping@Receiver}

The receiver in our case is an Android v$4.4.4$ ported Google Nexus~$5$ smartphone.
The complexity of the Android stack introduces several implementation challenges as the Bluetooth packet must propagate through its several layers before it can be received by the application layer. 
However, for accurate timing information, it is important that time stamping is done as close to the hardware (layer) as possible.
\newline
\indent
The reception time on the Android phone can be further divided into the following delivery delays:
\begin{enumerate}
	\item time taken by the Broadcom BCM radio to receive the message and raise an interrupt;
	\item time taken by the standard UART platform driver to hook into the BCM radio and register a RX event; and
	\item time taken by the Bluedroid stack to poll the UART driver, waiting to check if there are any bytes to be read (using the \texttt{userial}\_\texttt{read}\_\texttt{thread()} method in \texttt{userial.c}).
\end{enumerate}
\noindent
Therefore, there is a significant, nondeterministic time lag between the instants when a RF message is actually received by the radio to when it is processed by the Bluedroid stack.
Moving to software layers beyond the UART will limit portability across different Android phones.
Taking these facts into consideration, the lowest layer accessible entity on the Android Bluedroid stack is \texttt{userial.c}\footnote{In the Android Bluedroid stack, \texttt{userial.c} is located at: \texttt{/external/bluetooth/bluedroid/hci/src/}}.
\newline
\indent
\texttt{userial.c} interfaces with the standard UART driver, and provides a common interface for the time-sync to work on multiple phones without introducing many hardware changes.
However, timestamping at the level of \texttt{userial.c} introduces \emph{two} complications.
\emph{First}, it is a generic container for catching all types of events (notification and other messages) sent out by the BCM radio to the underlying driver, and hence, there is a need to correctly identify the Bluetooth event.
\emph{Second}, this problem can be overcome by waiting for \texttt{userial.c} to process the event list; but would introduce finite, variable delays.
Therefore, timestamps are recorded at the instant when an event is received at \texttt{userial.c}, but before any processing is performed on the list.
\newline
\indent
The \texttt{clock$\_$gettime(CLOCK$\_$REALTIME)} method, which returns the real-time clock of the system in nanoseconds since the Epoch ($00:00$ $1$ January, $1970$\,UTC), is utilized to perform low-level timestamping on the receiver (phone) end.
This representative value is passed on to the higher layer application, and the receive timestamp corresponding to the received BLE packet is subsequently determined.
It is possible that a notification was received by \texttt{userial.c} from the radio, and the application layer was called following that particular notification being recorded as a timestamp.
This introduces an error since the timestamp of the notification and not of the message is being used by the program.  
To overcome this problem, we adopt an algorithmic approach that determines the timestamp corresponding to a received packet.
The algorithm: first, iteratively records the timestamps for the previous and the current packet; and then, subtracts each current packet's timestamp value from every previous packet's timestamp value.
Upon completion, the value with the least deviation is taken as the best fit timestamp for the particular packet.

\subsection{Clock Drift Management}

Two clocks are considered synchronized when their relative deviation is bounded by some known value. 
However, as a result of clock drift arising from variations in clocking speed (frequency), two clocks will start to deviate even if they are set to identical values at initialization.
\newline
\indent
In our case, the clock quality and speed on the transmitter and the receiver are vastly different.
The nRF$24$\emph{Cheep} beacon unit has both high and low speed on-chip oscillators.
The high speed oscillator operates at the frequencies of $64$\,MHz, $48$\,MHz, $32$\,MHz, $16$\,MHz, $12$\,MHz, $8$\,MHz, $4$\,MHz, or $1$ MHz; while the low speed oscillator, which drives the timer, operates at $15$\,kHz.  
On the other hand, the Nexus~$5$ phone (control) unit is driven with a high precision clock pertaining to a few GHz.
The offset between these two clocks will change over time due to frequency differences in the oscillator, which may be short-term arising from environmental factors (such as temperature variations, change in supply voltage, etc.,), or long-term due to aging effects.
\emph{Cheep}Sync, rather than modeling the frequency instability, does continuous skew adjustments over a measurement window on the phone unit as: $k = (o_{i}-\overline{o})/(t_{i}^{r}-\overline{t^{r}})$; where, $\overline{t^{r}}$ is the average elapsed time at the reference clock and $\overline{o}$ is the average offset upto the $i^{th}$ sample point. 
This linear regression method offers a fast mechanism to find the frequency and phase errors over time.   
\newline
\indent
Fig.~\ref{fig:clockdrift} shows the performance of time synchronization with and without clock drift compensation for different $interval$ values, $100$\,ms and $200$\,$\mu$s, over a period of $13$\,hours.
Without drift compensation, the mean error in synchronizing the transmitter and receiver clock is in the \emph{millisecond} range, but can be bought down to a few \emph{microseconds} with correction (that includes both low-level timestamping and clock drift management).
For the specific case of $interval$ = $100$\,ms, the error before drift correction was: $\mu$=$2.62$\,ms, $\sigma^2$=$6.37312$\,$\mu$s and $95$\% cumulative probability of $8.2$\,ms; while the respective estimates after correction were recorded as: $\mu$=$0.0667$\,ms, $\sigma^2$=$3.89512$e-$02$\,$\mu$s and $0.64$\,ms at the $95$\% cumulative probability level (Fig.~\ref{fig:clockdrifta}).
Similar results were obtained for $interval$=$200$\,$\mu$s, where there was one order of  improvement in synchronization accuracy after compensating for clock drift, both in the average and $95$\% probability level (Fig.~\ref{fig:clockdriftb}).
The results also suggest that the accuracy and stability of the interval timer when set to $100$\,ms is significantly better than the case when it is configured to $200$\,$\mu$s.
The timestamp value when multiplied by $ts\_counter$ is more sensitive to changes in the interval timer because of its larger magnitude and error accumulation over time.
The interval timer of $100$\,ms was, therefore, chosen as the preferred $interval$ value due to its high stability, in addition to its lower energy cost. 

\subsection{Multi-device Time Synchronization}

Considering the system dynamics, there are \emph{two} forms of time synchronization across multiple devices.

\subsubsection{Many-Tx-to-One-Rx Synchronization}
The scenario of synchronizing multiple Tx's to a single Rx is a simple extension to the case of deriving an estimate of time for a single (transmitter, receiver) pair; wherein the control unit becomes the reference point to perform synchronization.
A reference point contains a pair of local and global timestamps where both of them refer to the same time instant.
The control unit receives periodic broadcasts from beacon units within their coverage zone; else, their records are not entered.
When the control unit collects the required measurement points, it estimates the skew and offset of the observed beacons and derives their coordinated time measure with respect to the global time.

\subsubsection{Many-Tx-to-Many-Rx Synchronization} \label{sec:many-tx-to-many-rx}
The scenario of synchronizing multiple Tx's to multiple Rx's combines the above system infrastructure with a mechanism for different control units to share the skew and offset information of already visited beacon units. 
This could be achieved by a peer-to-peer interaction between the control units or through the Cloud infrastructure.
For ease of implementation, we choose to take the later alternative with the Google Cloud platform.
In this case, the absolute timestamp of the control unit along with the timestamp of the beacon unit is inserted into the Cloud.
This data is then made accessible to other control units using the Google App Engine database. 
Whenever the timestamp of a new beacon unit is recorded by a control unit, a notification is sent to all other control units using Google cloud messaging.
Once a control unit receives this message, it observes the last time it obtained the timestamp of the same beacon and computes the time difference.
For instance, let the first control unit record timestamp $t_a$ from beacon $i$ at time instance $\tau_{a}$.
Let the second control unit record another timestamp from the same beacon $i$ with value $t_b$ at time instance $\tau_{b}$.
Using these information, the offset (i.e., time difference) between the phones is measured as: ($\tau_{a}-\tau_{b}-(t_b-t_a)*k$).
\newline
\indent
The relative time on the control units (that are Android phones) may vary from one to another by upto several seconds. 
It is, therefore, necessary to constantly synchronize with the nearest Network Time Protocol (NTP) server repeatedly in order to maintain very high precision for our time synchronization.
For our implementation, we used an Android NTP application\footnote{The Android NTP application is available at: \\\url{https://play.google.com/store/apps/details?id=ru.org.amip.ClockSync&hl=en}}, and was configured to synchronize the control units at an interval of $30$\,seconds.

\section{Experimental Evaluation} \label{sec:evaluation}
\begin{figure*}[p]
\begin{center}
\subfigure[many-tx-to-one-rx: $\mu$=$8\mu$s, CDF($95$\%)=$0.04$\,ms \vspace{-5mm}]{\label{fig:study1a}\includegraphics[width=3.3in]{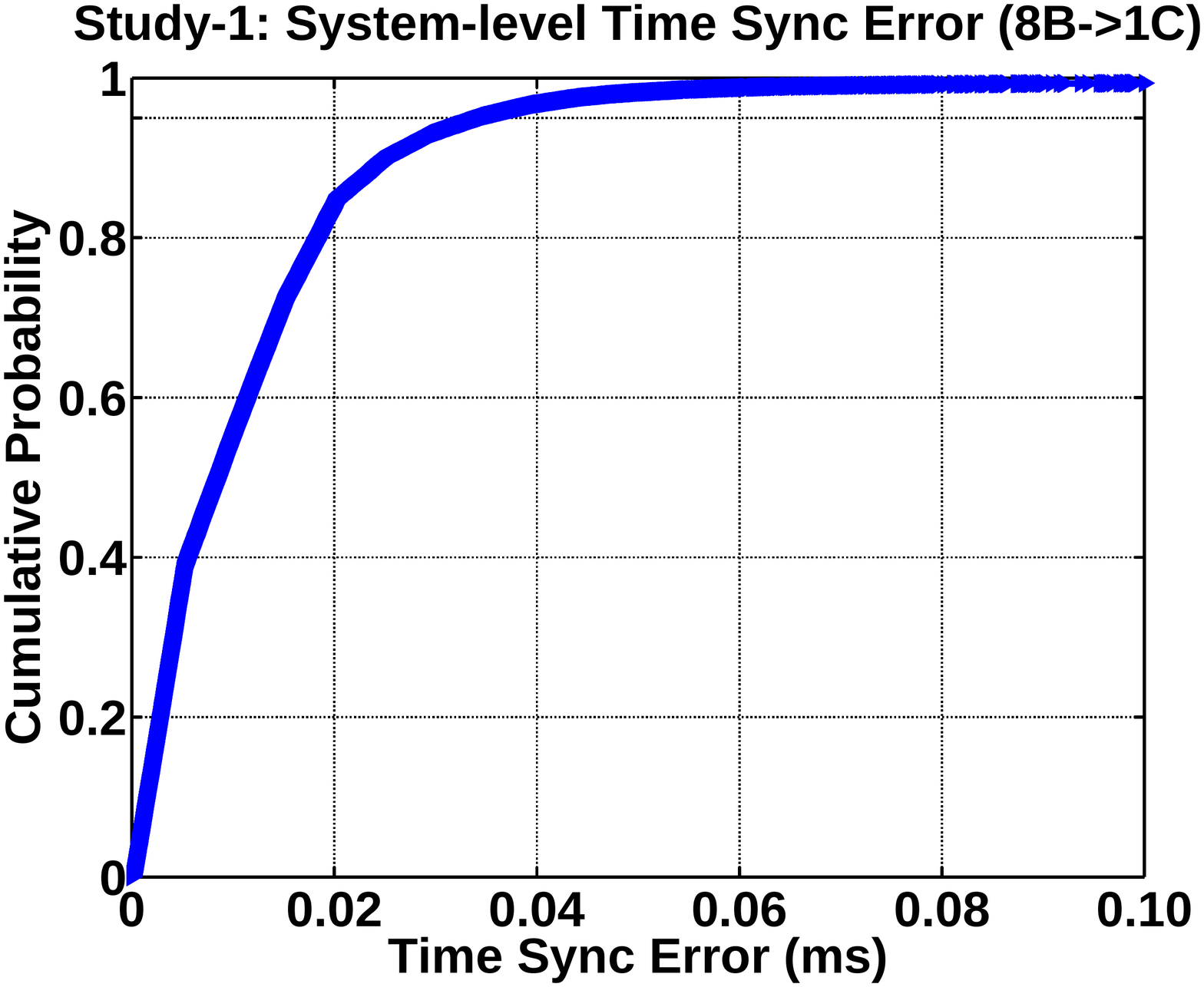}}
\subfigure[many-tx-to-one-rx\vspace{-5mm}]{\label{fig:study1b}\includegraphics[width=3.3in]{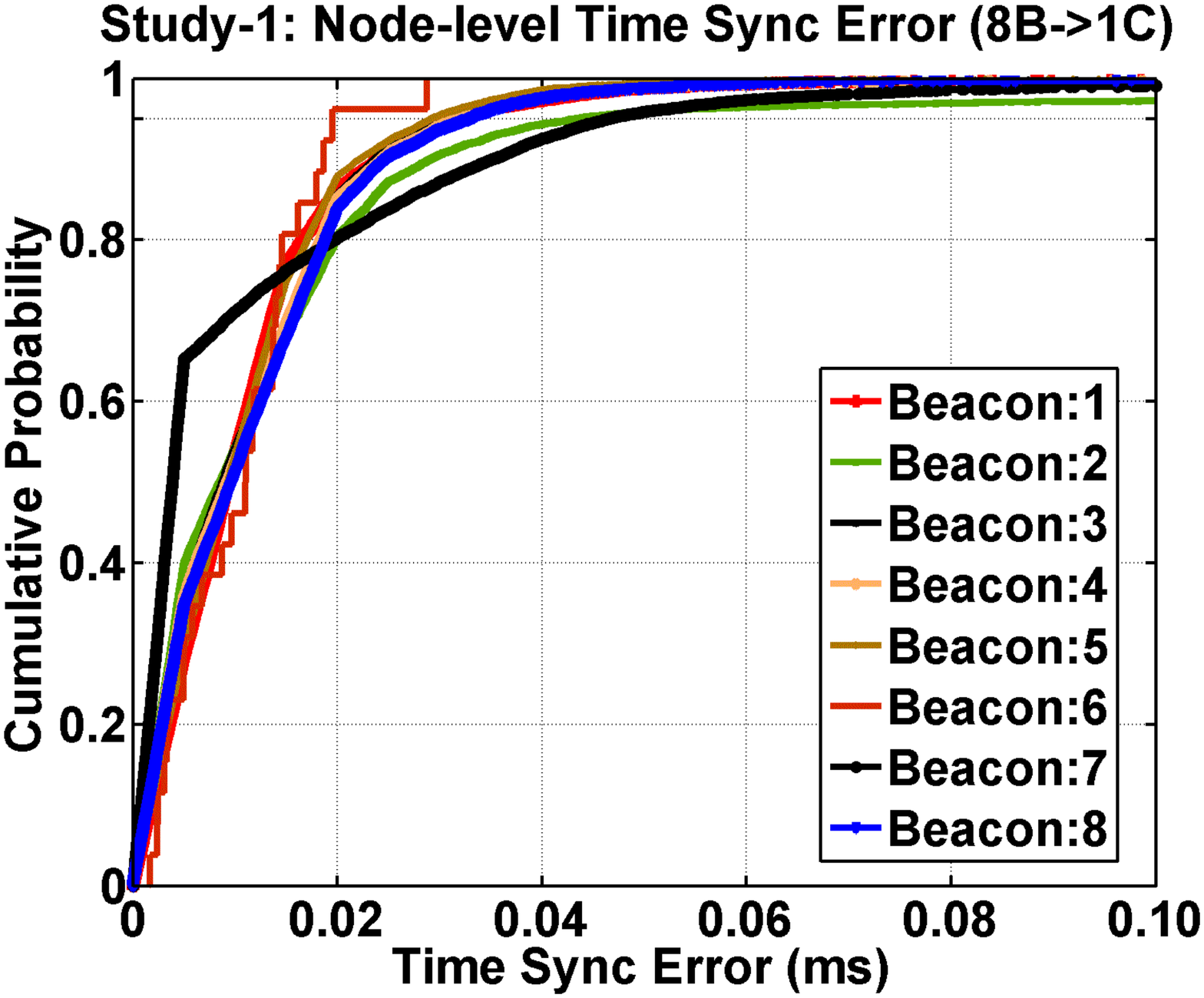}}
\subfigure[one-tx-to-many-rx scenario: $\mu$=$10\mu$s, CDF($95$\%)=$5$\,ms]{\label{fig:study1c}\includegraphics[width=3.3in]{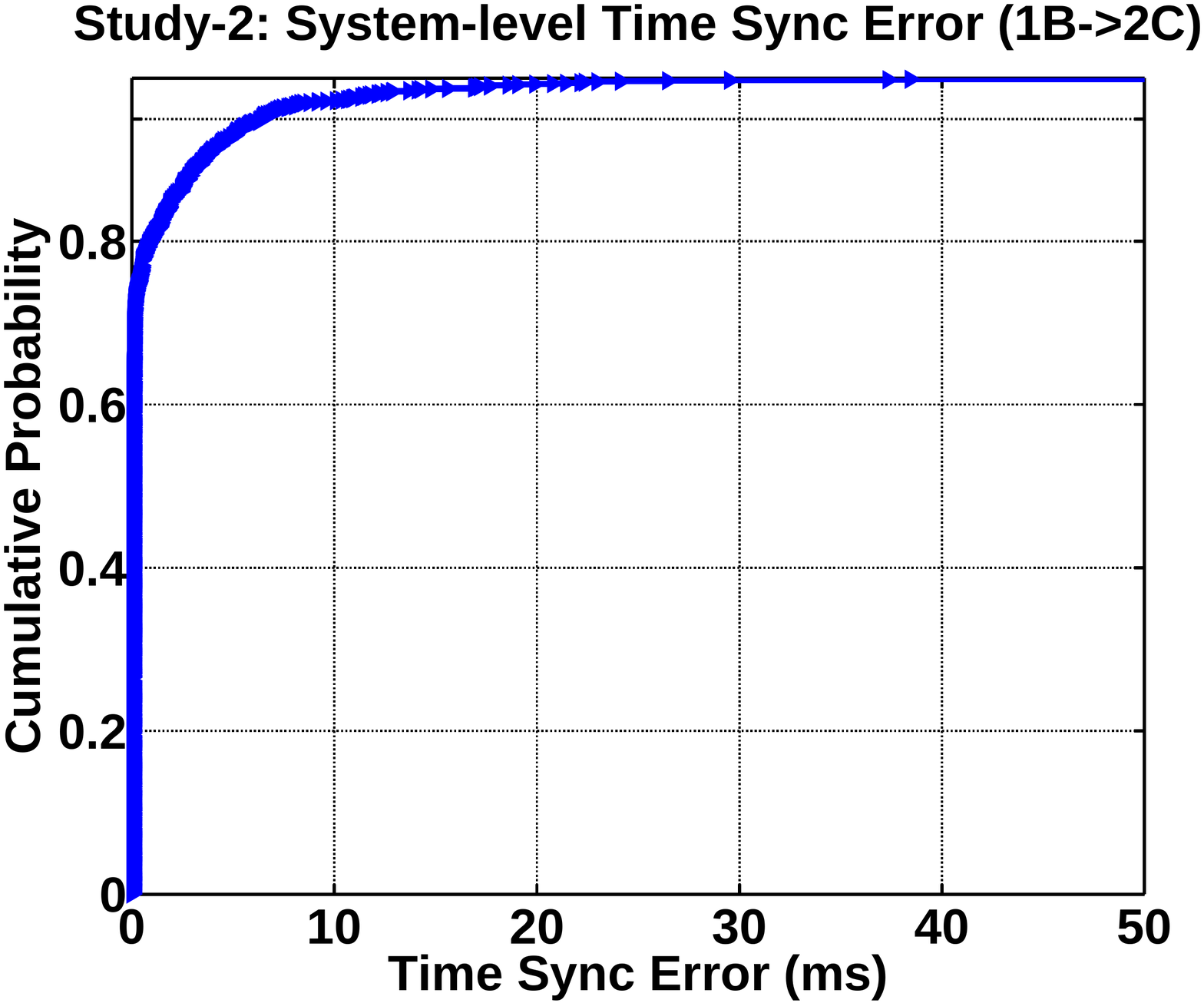}}\hspace{0.7cm}
\subfigure[many-tx-to-many-rx scenario: $\mu$=$10\mu$s, CDF($95$\%)=$10$\,ms]{\label{fig:study1d}\includegraphics[width=3.3in]{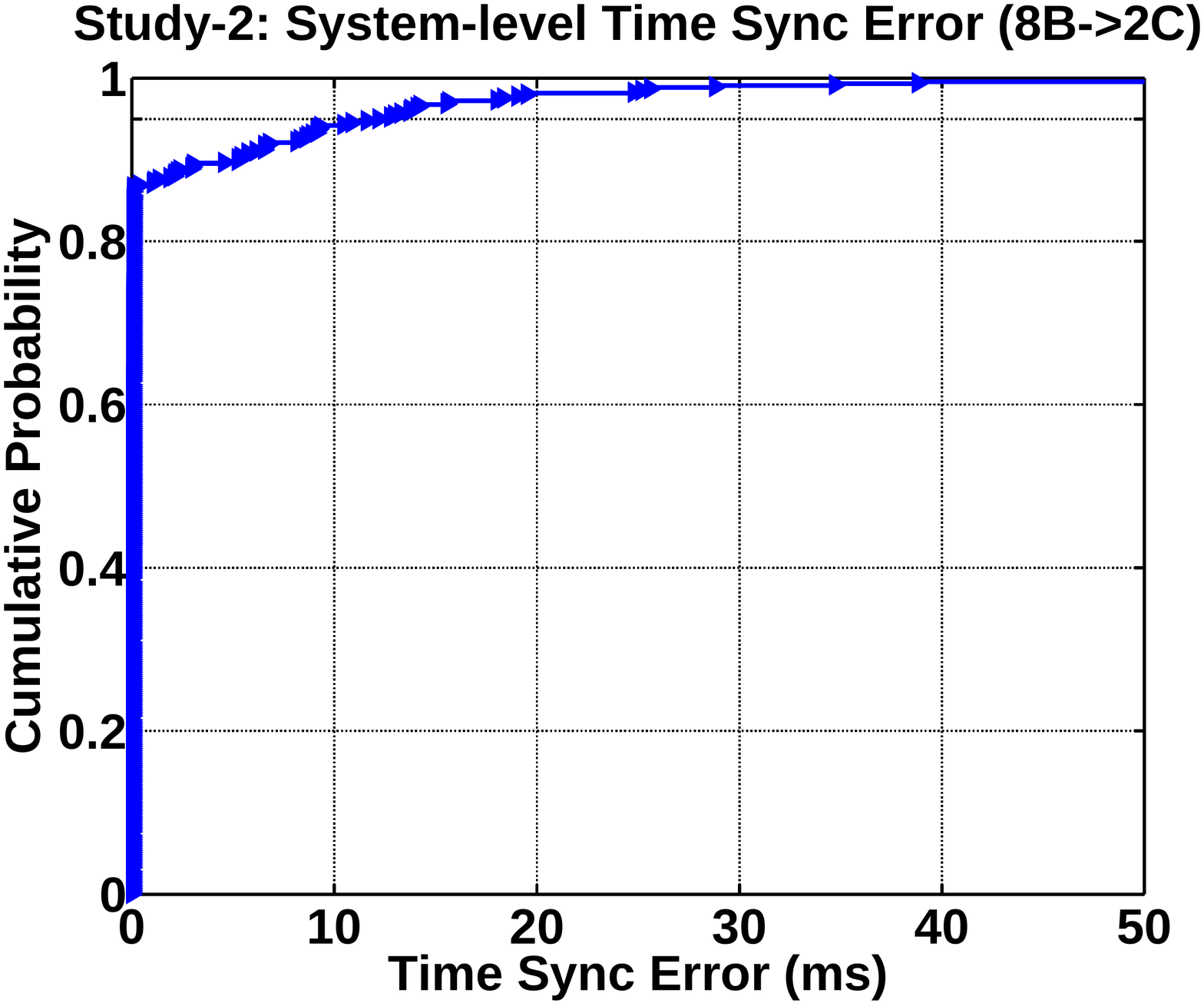}}
\subfigure[many-tx-to-one-rx scenario: $\mu$=$12\mu$s, CDF($95$\%)=$0.04$\,ms]{\label{fig:study1e}\includegraphics[width=3.3in]{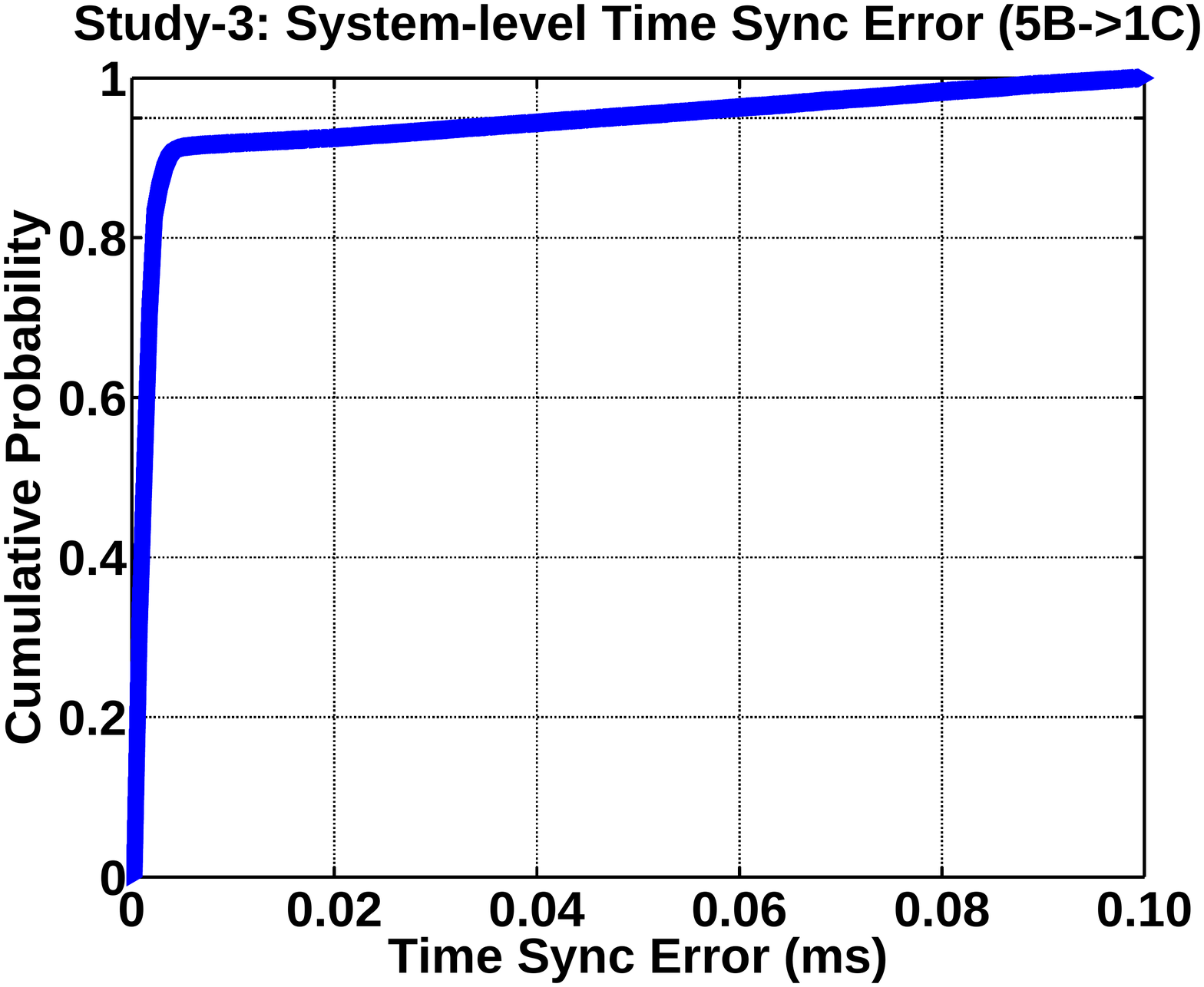}}\hspace{0.7cm}
\subfigure[many-tx-to-many-rx scenario: $\mu$=$10\mu$s, CDF($95$\%)=$22$\,ms]{\label{fig:study1f}\includegraphics[width=3.3in]{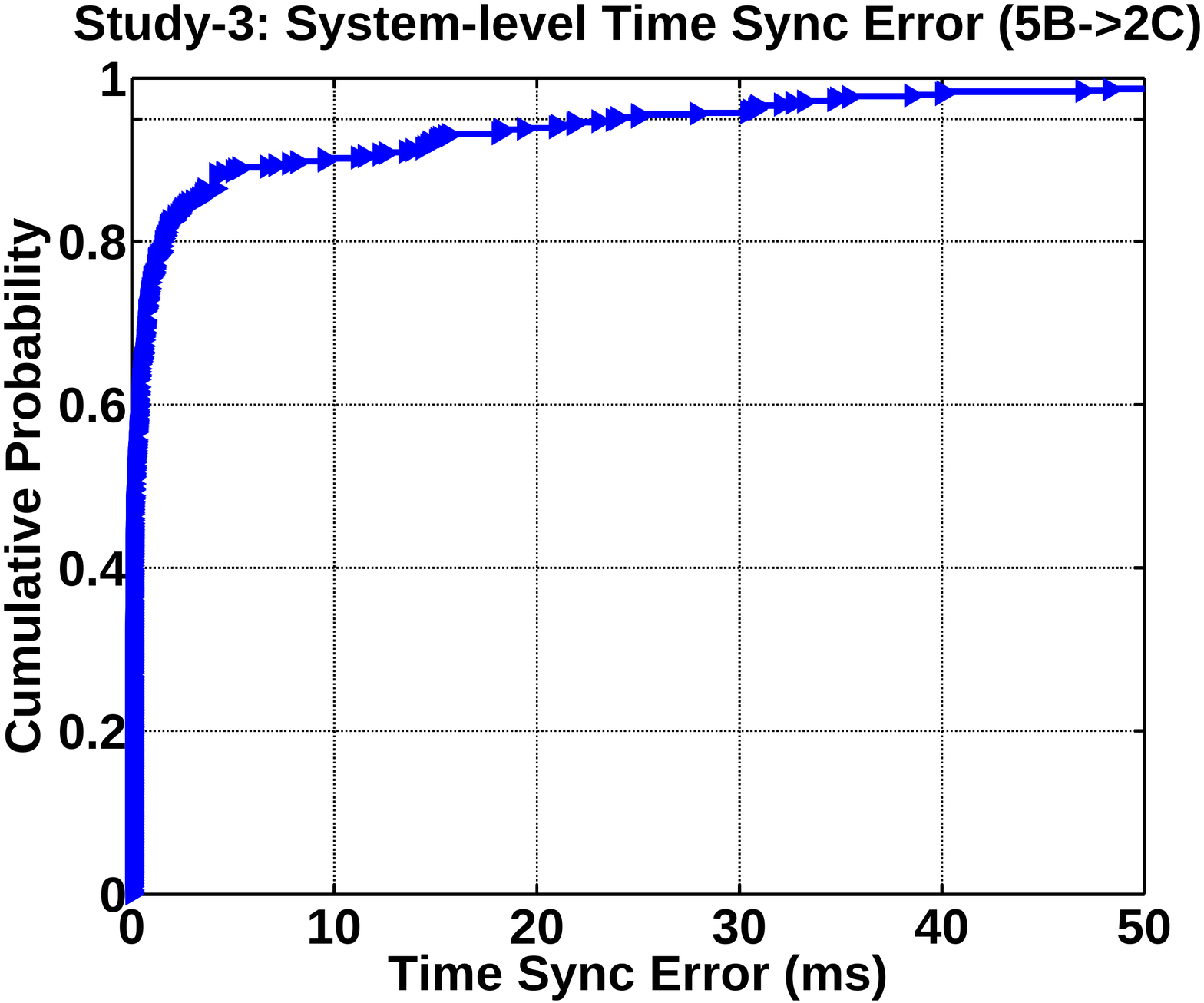}}
\end{center}
\caption{\textbf{Performance of \emph{Cheep}Sync.}}
\label{fig:study1}
\end{figure*}

In this section, we evaluate the accuracy of \emph{Cheep}Sync in a variety of controlled and uncontrolled experiments.
The controlled experiments provide benchmark results for validating against uncontrolled experiments.

\subsection{CheepSync in Controlled Experiments}
\vspace{1mm}
\noindent
\textbf{Study-$1$.} The aim of this study was to obtain the baseline performance level of CheepSync for a \emph{many-tx-to-one-Rx} synchronization scenario.
It was, therefore, designed with a \emph{static} setup of $8$ beacon units and $1$ control unit that was always covered by all the beacons.
The beacons were configured to broadcast advertisement packets at an interval of $100$\,ms and at their lowest transmit power of $-20$\,dB. 
This experiment was conducted for about $13$\,hours wherein an average of $10000$ packets were received by the control unit from each beacon.
Here, it should be noted that communication between the beacon and control units was over an interference channel as BLE does not provision for medium access control rules in the broadcast, advertisement mode.
\newline
\indent
The experimentation results are shown in Fig.~\ref{fig:study1}.
The system-level performance of \emph{Cheep}Sync is depicted in Fig.~\ref{fig:study1a}, which shows an average error is $8$\,$\mu$s and a $95\%$ error probability of less than $0.04$\,ms.
Fig.~\ref{fig:study1b}, disaggregates the combined (drift compensated) measurements into respective representations for each of the $8$ beacons in the system.
Here, it is evident that a large percentage of the beacon units show consistent behavior with an average error level of approximately $10$\,$\mu$s and a worst case value of $0.04$\,ms; except beacon~$7$ that shows a $2$~times better average performance but shoots over in the worse case performance level.  
\vspace{2mm}
\newline
\noindent
\textbf{Study-$2$.} The aim of this study was to obtain the baseline performance level of \emph{Cheep}Sync for the scenarios of \emph{one-tx-to-many-rx} and \emph{many-tx-to-many-rx} synchronization. 
For the \emph{prior} scenario, the static setup consisted of a single beacon and $2$ control units; while for the \emph{later} scenario, the setup consisted of $8$ static beacons and $2$ static control units.
The control units were always kept within the coverage range of the beacon(s); while all other beacon/control unit configuration parameters were kept consistent with study-$1$.
Fig.~\ref{fig:study1c} and Fig.~\ref{fig:study1d} shows the cumulative time-sync error (i.e., the time difference between control units as explained in Section~\ref{sec:many-tx-to-many-rx}) distributions for the prior and later setup respectively, and it is conditioned on contributing errors factors such as clock drift, NTP drift, etc.,.
The result obtained from Fig.~\ref{fig:study1c} suggests that there is a $95$\% probability of the time synchronization error to be less than $5$\,ms with mean $< 10\mu$s.
A similar observation is also noted from Fig.~\ref{fig:study1d} where the result suggests that there is a $90$\% probability of the time sync error to be less than $10$\,$\mu$s; although with the possibility of errors as large as $10$\,ms for $5$\% higher confidence levels. 

\subsection{CheepSync in Uncontrolled Experiments}
\vspace{1mm}
\noindent
\textbf{Study-$3$.} This experiment was designed to evaluate the performance of \emph{Cheep}Sync in an uncontrolled setup, and hence, quantify its deviation from the respective benchmark results obtained from study-$1$ and study-$2$.
The experimentation space was in a [$20$$\times$$20$]\,m portion of our office floor.
In a real-world setting, we expect our office staff to be running \emph{Cheep}Sync service on their smartphones as they walk through the various sections of this office space.
They are also unlikely to be continuously walking as they would walk between locations of interest and spend more time at certain places.
For our evaluation, we took the services of two people who were handed an Android smartphone running the \emph{Cheep}Sync time service and they kept it with themselves for about an hour.
The respective office space was instrumented with a set of $5$ beacon units in a manner that all beacons did not cover every end of the experiment zone.
All other system configurations such as the beaconing rate and transmit power levels were kept the same as in study-$1$/$2$.
\newline
\indent
For \emph{many-tx-to-one-rx} scenario (Fig.~\ref{fig:study1e}), a mean error of $12$\,$\mu$s was observed while its $95$\% error probability was less than $0.04$\,ms.
As for the \emph{many-tx-to-many-rx} scenario (Fig.~\ref{fig:study1f}), the $95$\% error probability was less than $22$\,ms with a mean error of $10$\,$\mu$s.
Both of these results are in good agreement with the baseline observations recorded in Fig.~\ref{fig:study1a} and Fig.~\ref{fig:study1c}; thereby establishing the efficacy of \emph{Cheep}Sync.

\section{Related Work} \label{sec:related_work}
 
Time synchronization has different perceptions. 
The first thought is to create a right chronology of events in a network by \emph{ordering}.
Lamport showed that the knowledge of the exact time instants is not required for achieving this weak notion of synchronization; and that a total order to events can be obtained by virtual clocks\cite{Lamport1978:TCO}.
The other perceptions are closer to the stronger sense of synchronization, and it is to derive a common time reference across distributed systems in a \emph{relative} or \emph{absolute} manner. 
\newline
\indent
While there exist many solutions in this space, they have some basic features in common: a messaging protocol to exchange time-stamped packets between nodes (some acting as client and others as time servers) in a network, techniques for overcoming nondeterministic delays, and an adjustment mechanism to update the local clock.
However, they differ in aspects such as: whether the physical clocks of the network are kept consistent internally, or are synchronized to external standards; if the server is an arbiter of the client clock or is considered as a canonical clock, etc.,. 
\newline
\indent
For synchronization in WSNs, there exist multiple algorithms that use messaging passing as the basic mechanism to compensate for delays. 
The Reference Broadcast Synchronization (RBS)\cite{Elson2002:RBS} scheme, for example, uses packet broadcasts and exchange of time information among multiple receivers to eliminate transmission delays.
By averaging over multiple transmissions, this method is able to achieve tight pairwise clock synchronization and reduces timing jitters associated with embedded devices.
The Flooding Time Synchronization Protocol (FTSP)\cite{Maroti2004:FTSP} and Time-sync Protocol for Sensor Networks (TPSN) \cite{Ganeriwal2003:TPSN} use low-level hardware timestamping to eliminate similar time jitters.
While both techniques use message broadcasts to form a spanning tree of all nodes in the network; FTSP (unlike TPSN) periodically adjusts the local clock rate to compensate for drift and also handles dynamic topology changes.
Using flooding as the basic communication primitive (like FTSP), Glossy\cite{Ferrari2011:glossy} uses a novel architecture that uses constructive interference to its advantage and implicitly synchronizes nodes as the flooding packet propagates through the network.
Compared to  its processors, Glossy delivers higher accuracy by flooding packets within a few milliseconds using the approach of Virtual High-resolution Time (VHT). 
Rowe \emph{et al.} \cite{Rowe2009:LCS} introduced hardware assisted clock synchronization in WSNs.
Here the key idea was to tune into the magnetic field radiating from existing AC power lines, and thus leverage a highly available common clock source for synchronization.
\newline
\indent
\emph{Cheep}Sync explores a form of synchronization that differs from traditional WSNs.
Its fundamental design philosophy is to provision for synchronization between transmitters and receivers, as opposed to traditional WSN protocols that synchronize a set of receivers with one another.
Performing receiver-receiver synchronization removes uncertainties associated with send and access delays, the biggest contributors to non-determinism in latency, from the critical time path. 
System level complexities are further reduced by using symmetric WSN platforms that run on simple OS.
All of these factor conglomerates towards higher levels of synchronization accuracy that is typically evident in traditional WSNs.
\emph{Cheep}Sync, in contrast, operates on highly asymmetric devices with vast differences in system complexity.
Its synchronization methodology, system architecture, and wireless communication standard utilized for delivering the respective solutions are completely different.
\emph{Cheep}Sync achieves a high synchronization accuracy that is better than the time-sync levels of RBS, TPSN and Rowe et al.; but is less precise compared to FTSP or Glossy (Table~\ref{tab:timesync}).

\begin{table}[t]
  \begin{center}
  \caption{\textbf{Accuracy Measure of Time Synchronization Algorithms}}
	\begin{tabular}[c]{lr}
	\hline
		{\bf Time Sync. Protocol} & {\bf Avg. Accuracy ($\mu$s)} \\
		 & {\bf (Single Hop)} \\
		\hline
		RBS \cite{Elson2002:RBS} & $29.10$ \\
		TPSN \cite{Ganeriwal2003:TPSN} & $16.90$ \\
		FTSP \cite{Maroti2004:FTSP} & $01.48$ \\
		Glossy \cite{Ferrari2011:glossy} & $0.50$ \\
		Rowe et al. \cite{Rowe2009:LCS} & $1000.00$ \\
		\emph{Cheep}Sync  & $10.00$ \\
		\hline
\label{tab:timesync}
\end{tabular}
\end{center}
\end{table}
\section{Conclusion} \label{sec:conclusion}

\emph{Cheep}Sync is a time synchronization service for BLE advertisers; and therefore, is a key enabler for a variety of IoTH applications where mobile crowdsourcing, using existing infrastructure and ubiquitously used platforms along with `humans' as the data mule, has emerged as an alternate architecture.
These applications typically require different levels of time accuracy.
For instance, a few milliseconds may be sufficient to disambiguate the handwash event of medical practitioners in a hospital setting, but a significantly higher accuracy level of a few microseconds would be required to secure the same system against possible gaming.
By empirical evaluations, we showed that \emph{Cheep}Sync is capable of gracefully handling timing requirements as low as $10$\,$\mu$s.
It is build on the existing Bluetooth v$4.0$ standard, and hence is, generic to all devices using the low energy profile of Bluetooth.

\bibliographystyle{unsrt}
\bibliography{cheepsync-references} 

\begin{thebibliography}{10}

\bibitem{Lamport1978:TCO}
L.~Lamport.
\newblock Time, clocks, and the ordering of events in a distributed system.
\newblock {\em Communications of the ACM}, 21(7):558--565, July 1978.

\bibitem{Maroti2004:FTSP}
M.~Mar\'{o}ti, B.~Kusy, G.~Simon, and \'{A}. L{\'e}deczi.
\newblock The flooding time synchronization protocol.
\newblock In {\em Proceedings of the 2nd International Conference on Embedded
  Networked Sensor Systems}, SenSys '04, pages 39--49, New York, NY, USA, 2004.
  ACM.

\bibitem{Elson2002:RBS}
J.~Elson, L.~Girod, and D.~Estrin.
\newblock Fine-grained network time synchronization using reference broadcasts.
\newblock In {\em Proceedings of the 5th Symposium on Operating Systems Design
  and Implementation}, OSDI '02, pages 147--163, New York, NY, USA, 2002. ACM.

\bibitem{Ganeriwal2003:TPSN}
S.~Ganeriwal, R.~Kumar, and M.B. Srivastava.
\newblock Timing-sync protocol for sensor networks.
\newblock In {\em Proceedings of the 1st International Conference on Embedded
  Networked Sensor Systems}, SenSys '03, pages 138--149, New York, NY, USA,
  2003. ACM.

\bibitem{Ferrari2011:glossy}
F.~Ferrari, M.~Zimmerling, L.~Thiele, and O.~Saukh.
\newblock Efficient network flooding and time synchronization with glossy.
\newblock In {\em Proceedings of the 10th International Conference on
  Information Processing in Sensor Networks}, IPSN '11, pages 73--84, April
  2011.

\bibitem{Mills1991:NTP}
D.L. Mills.
\newblock Internet time synchronization: the network time protocol.
\newblock {\em IEEE Transactions on Communications}, 39(10):1482--1493, Oct
  1991.

\bibitem{Gusella1989:tempo}
R.~Gusella and S.~Zatti.
\newblock The accuracy of the clock synchronization achieved by tempo in
  berkeley unix 4.3bsd.
\newblock {\em IEEE Transactions on Software Engineering}, 15(7):847--853, Jul
  1989.

\bibitem{Cristian1989:probtimesync}
F.~Cristian.
\newblock Probabilistic clock synchronization.
\newblock {\em Distributed Computing}, 3(3):146--158, 1989.

\bibitem{Kopetz1987:CSD}
H.~Kopetz and W.~Ochsenreiter.
\newblock Clock synchronization in distributed real-time systems.
\newblock {\em IEEE Transactions Computers}, 36(8):933--940, August 1987.

\bibitem{Rowe2009:LCS}
A.~Rowe, V.~Gupta, and R.~Rajkumar.
\newblock Low-power clock synchronization using electromagnetic energy
  radiating from ac power lines.
\newblock In {\em Proceedings of the 7th ACM Conference on Embedded Networked
  Sensor Systems}, SenSys '09, pages 211--224, New York, NY, USA, 2009. ACM.

\bibitem{BLE}
{Specification Adopted Documents}.
\newblock
  \url{https://www.bluetooth.org/en-us/specification/adopted-specifications}.

\bibitem{RenesasRL78G14}
{Renesas RL78/G14}.
\newblock
  \url{http://www.am.renesas.com/products/mpumcu/rl78/rl78g1x/rl78g14/index.jsp}.

\bibitem{nRF2401}
{nRF2401+ Ultra Low power 2.4GHz RF Transceiver}.
\newblock \url{http://www.nordicsemi.com/eng/Products/2.4GHz-RF/nRF24L01P}.

\bibitem{ble-fakery}
{Bit-Banging Bluetooth Low Energy}.
\newblock
  \url{http://dmitry.gr/index.php?r=05.Projects&proj=11.%20Bluetooth%20LE%20fakery}.

\bibitem{Horauer2001:psync}
M.~Horauer, K.~Schossmaier, U.~Schmid, and N.~Kero.
\newblock Psynutc—evaluation of a high precision time synchronization
  prototype system for ethernet lans.
\newblock In {\em Proceedings of the 34th Annual Precise Time and Time Interval
  Meeting}, PTTI '02, pages 263--279, 2002.

\end{thebibliography}

\end{document}